\documentclass[aip,apl,reprint,showpacs]{revtex4-1}
\usepackage{graphicx}

\begin{document}


\title{Charge Detection in Phosphorus-doped Silicon Double Quantum Dots} 


\affiliation{Hitachi Cambridge Laboratory, J.J. Thomson Avenue, Cambridge, U.K.}
\author{A. Rossi}
\email[Electronic mail: ]{ar446@cam.ac.uk}
\author{T. Ferrus}
\author{G.J. Podd}
\author{D.A. Williams}


\date{\today}

\begin{abstract}
We report charge detection in degenerately phosphorus-doped silicon double quantum dots (DQD) electrically connected to an electron reservoir. The sensing device is a single electron transistor (SET) patterned in close proximity to the DQD. Measurements performed at 4.2K show step-like behaviour and shifts of the Coulomb Blockade oscillations in the detector's current as the reservoir's potential is swept. By means of a classical capacitance model, we demonstrate that the observed features can be used to detect single-electron tunnelling from, to and within the DQD, as well as to reveal the DQD charge occupancy.
\end{abstract}


\maketitle 

Silicon-based systems for quantum computing have attracted much attention because of their scalability, their well-established technological process and their long coherence time.\cite{fujisawa}
Both intrinsic and extrinsic silicon substrates have been widely employed for the fabrication of devices aiming for single charge detection.\cite{angus,single} In particular, Si:P single electron transistors (SET) capacitively coupled to isolated double quantum dots\cite{idqd} have provided a route for the implementation of charge qubits.\cite{gorman} The electrical isolation of this type of qubit results in significant suppression of decoherence due to the decoupling of the electronic states from the leads. On the other hand, the correct operation of these devices strongly depends on the density of excess charge permanently stored in the isolated structure as a result of the fabrication process. This may lead to limited system reliability and process yield.\\\indent
\begin{figure}[b!]
\includegraphics[scale=0.34]{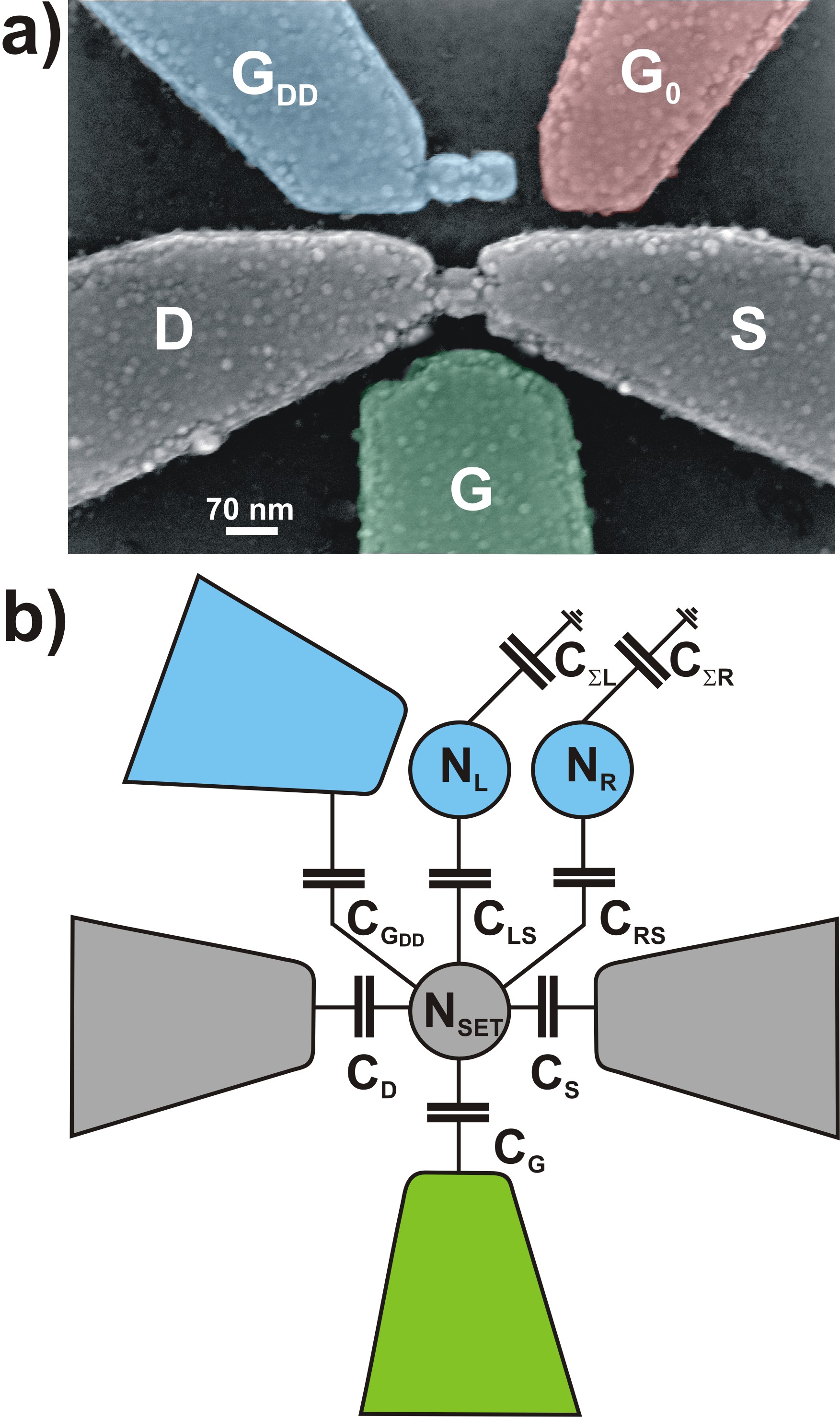}
 \caption{\label{fig:layout}(Color online) a) False color scanning electron micrograph of a device similar to those investigated. Lower structure defines the SET detector which is made up of Drain (D), Source (S) and Gate (G). Upper part shows the DQD connected to the lead (G$_\mathsf{DD}$) and an additional control gate (G$_\mathsf{0}$). b) Equivalent circuit including the SET dot coupling capacitances and the total capacitances of the DQD left and right dots. Coupling to gate G$_\mathsf{0}$ is not shown for simplicity.}
 \end{figure}
In order to assess the impact of different electron densities on the dots' behavior, we have realised a Si:P double quantum dot (DQD) electrically connected to an electron reservoir via a tunnel barrier. Unlike previously reported work, in this system the charge density is bias-dependent and one can make electrons tunnel in, out or within the DQD by simply changing the voltage of a control electrode. The charge state of the DQD is detected by a single electron transistor patterned in close proximity and made of the same material.\cite{augke,mizuta} Distinctive features in the Coulomb Blockade (CB) characteristic of the SET allow us to demonstrate that single electron transfer and charge occupancy in the DQD can be readily sensed by the capacitively coupled detector device. \\\indent
The layout of a device similar to those investigated is shown in Fig.~\ref{fig:layout}(a). The base material used is silicon-on-insulator (SOI) with a 35 nm thick phosphorus-doped active layer (donor concentration $\sim$~3$\times$10$^{19}$~cm$^{-3}$). Full details of the system material and fabrication process are reported elsewhere.\cite{fab} Quantum confinement of excess electrons in each dot is achieved by patterning constrictions in the SOI nano-structure which act as tunnel barriers due to carrier depletion by sidewall trapping.\cite{mypaper} Although the gate electrodes G and G$_\mathsf{0}$ are meant to control the electrochemical potential of the SET and DQD respectively, each gate is capacitively coupled to both systems and will affect them to a different extent according to geometrical considerations. Similarly, the gate electrode G$_\mathsf{DD}$, apart from prominently controlling tunnelling events in the DQD, will also influence the SET potential, as depicted in the capacitance model of Fig.~\ref{fig:layout}(b).
\begin{figure}[t]
\includegraphics[scale=0.70]{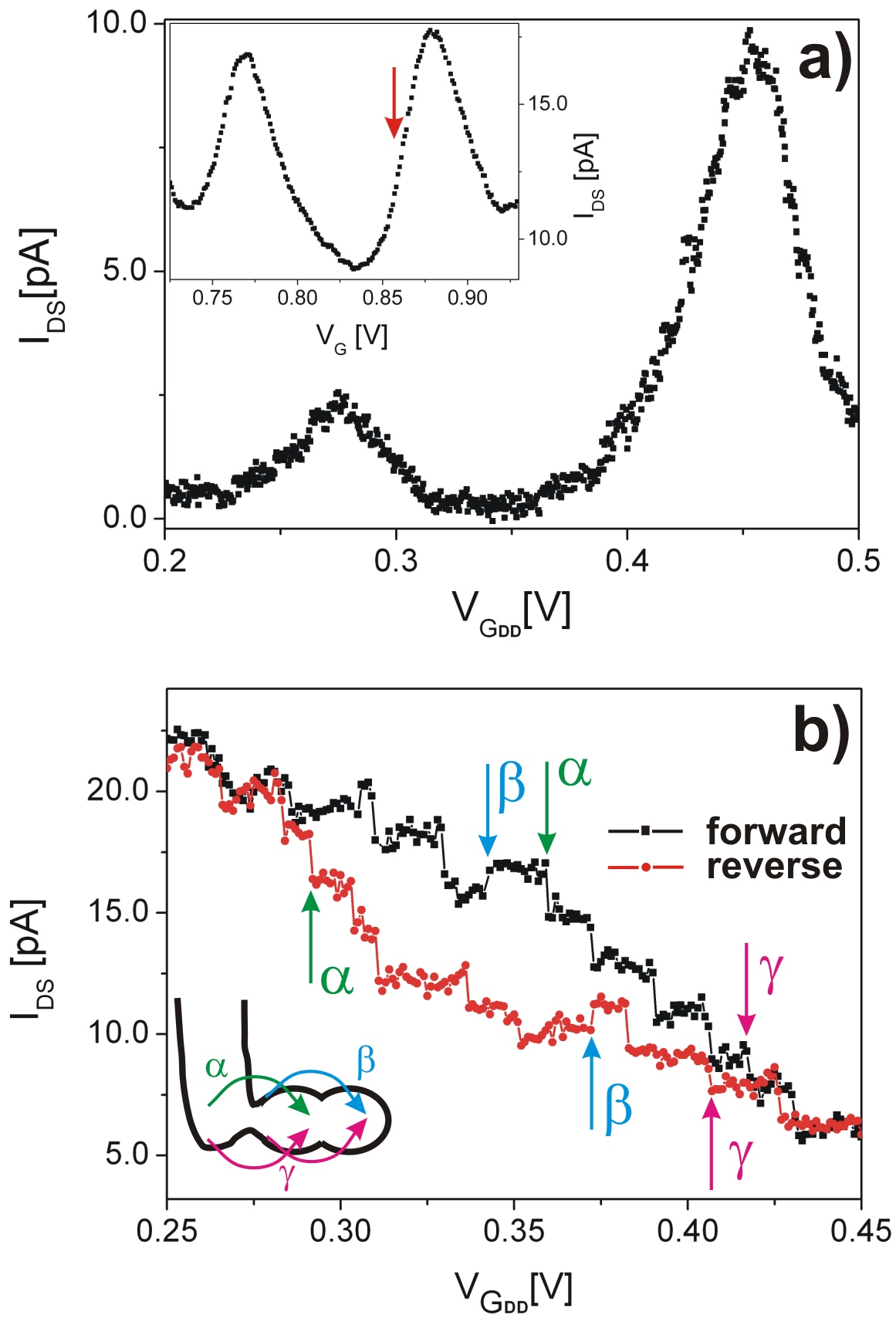}
\caption{\label{fig:data}(Color online) Detector's current response. a) CB oscillations induced by sweeping the DQD gate voltage. In absence of compensation, G$_\mathsf{DD}$ acts as a gate for the SET and can modify its bias point. Inset: CB oscillations as a function of $V_G$. The arrow indicates the bias point chosen to maximize charge sensitivity ($\sim$250pA/V). b) SET current as $V_{G_{DD}}$  is swept in forward (black squares) and reverse (red dots) direction. $V_{DS}$=30mV, $V_{G_0}$=0V; $V_G$ is swept with a compensation ratio of -2/5 to $V_{G_{DD}}$ and is centered at the operating point illustrated in the inset of (a). Step-like behavior indicates the occurrence of single-electron transfers. Arrows of different color and lettering highlight steps produced by different kinds of tunnelling events. Inset: cartoon sketching DQD tunneling associated to each family of current steps.}
\end{figure}
From the layout of the device is also clear that the left dot of the DQD is more strongly coupled to the SET island than the right dot. All experiments reported here have been carried out by directly immersing the samples into liquid helium at 4.2K.	\\\indent
In the experiments, the detector's operating point is set on the steepest part of the slope of a CB oscillation (see inset of Fig.~\ref{fig:data}(a)), so that the SET would work in the most charge-sensitive regime. In order to trigger tunneling events in the DQD, the voltage of the electrode G$_\mathsf{DD}$ must be swept. However, for the reasons mentioned above, a modification of $V_{G_{DD}}$ would inevitably have a secondary effect and move the SET operating point away from the selected range. Fig.~\ref{fig:data}(a) reveals that the SET current undergoes nearly two complete CB oscillations when $V_{G_{DD}}$ is swept from 0.2V to 0.5V. In order to minimize this unwanted shift in the bias condition, the SET gate voltage ($V_G$) is simultaneously swept with an appropriate compensation ratio which maintains the SET near the selected operating point. The detector's response to a $V_{G_{DD}}$ compensated sweep is reported in Fig.~\ref{fig:data}(b). We clearly see discrete jumps in the drain-source current. The step-like behavior is completely reproducible over many cycles despite the presence of some hysteretic effect, as it can be seen from the two subsequent voltage sweeps in opposite direction reported. Defects at the interface between Si and SiO$_\mathsf{2}$ acting as trap states are known to be responsible for charge offset drift in this kind of devices.\cite{zimm_offset} Therefore, it is not surprising that the two traces do not completely overlap. Multiple gate voltage preliminary sweeps turned out to be effective in neutralising the majority of these defect states and improve reproducibility. Discretization of the current of a SET capacitively coupled to a quantum dot is the evidence of detection of single electron tunneling to/from the dot.\cite{stone} The height and the relative direction of each step provide information on the nature of the tunneling event. We have identified three different families of steps whose amplitudes are approximately +1pA, -1pA and -2pA; two examples of each type are highlighted in the figure with blue ($\beta$), pink ($\gamma$) and green ($\alpha$) arrows, respectively. Plus signs indicate that current increases (decreases) when the voltage is increased (decreased), minus signs indicate that current increases (decreases) when the voltage is decreased (increased). As sketched in the inset of Fig.~\ref{fig:data}(b), steps of -2pA in height are interpreted as the transfer of one electron from the reservoir to the left dot, +1pA steps as a transfer of one electron from the left to the right dot, steps of -1pA are due to the two previous events taking place simultaneously. Interestingly, this latter event would produce an effect which is the linear combination of the previous ones (i.e. -2pA+1pA=-1pA), as we shall discuss later.\\\indent  Besides discrete changes in the detector's current level, it has been demonstrated that a single charging event in a quantum dot also results in a shift in the position of the conductance peaks.\cite{hofmann,gareth} In Fig.~\ref{fig:levels}(a) we show the detector's CB oscillations as the SET gate (G) is swept for different voltages applied to the DQD gate (for this type of experiment $V_{G_0}$ is used to compensate the effect of $V_{G_{DD}}$ on the SET). The plot shows that peaks can be shifted in either direction with respect to the SET gate voltage; these shifts can be associated to the current steps of Fig.~\ref{fig:data}(b), as the color and the lettering of the arrows suggest. In fact, shifts are in the range 4mV to 8mV which corresponds to 3\% to 6\% of the observed CB oscillations period ($\sim$130mV). This is in agreement with the expected voltage shift that can be evaluated from current steps of 1pA and 2pA at the selected SET operating point ($\sim$250pA/V).\\\indent
In order to model the effects of single charging in the DQD, we use an equivalent circuit representation of the system. Every device element (gate, dot, lead) is coupled to every other element via a coupling capacitance. In Fig.~\ref{fig:layout}(b), we only report the SET dot coupling capacitances, whereas the DQD coupling capacitances are embedded in the total capacitance terms, $C_{\Sigma_R}$ and $C_{\Sigma_L}$, for right and left dot, respectively. The addition of a single electron to either dot would discretely modify the potential of the SET dot proportionally to the specific inter-dot coupling. This would give rise to the observed current steps and CB shifts. However, since left and right dot are not equally coupled to the SET island (i.e. $C_{RS}\neq C_{LS}$), its electrochemical potential would be affected to a different extent according to which side of the DQD an electron is transferred to. Simple calculations of the SET dot energy demonstrate how current steps/shifts of different magnitude and direction can arise due to this coupling mismatch. We can evaluate the electrostatic energy of the SET island as a function of the excess electron number on each quantum dot as:
\begin{equation}
E=\frac{e^2N_{SET}^2}{2C_\Sigma}+\frac{e^2N_{SET}}{C_\Sigma}(\frac{C_{LS}}{C_{\Sigma_ L}}N_L+\frac{C_{RS}}{C_{\Sigma_R}}N_R)
\end{equation}
being $C_\Sigma$ the SET dot total capacitance and $N_{SET}$, $N_L$, $N_R$ the excess numbers of electrons in the SET island, DQD left dot and DQD right dot, respectively. The experimental condition of gate voltage compensation is taken into account by keeping constant the value of $N_{SET}$, whereas an electron tunneling to or within the DQD would modify the value of either $N_L$ or $N_R$. In particular, tunneling from the reservoir to the left dot would increase $N_L$ by one and change the SET dot energy by an amount:
\begin{equation}
\Delta_\alpha=\frac{e^2N_{SET}}{C_\Sigma}\frac{C_{LS}}{C_{\Sigma_L}}
\end{equation}
Tunneling from the left to the right dot would decrease $N_L$ by one, increase $N_R$ by one and produce an energy shift
\begin{equation}
\Delta_\beta=\frac{e^2N_{SET}}{C_\Sigma}(\frac{C_{RS}}{C_{\Sigma_R}}-\frac{C_{LS}}{C_{\Sigma_L}})
\end{equation}
The simultaneous tunneling events would leave $N_L$ unchanged, increase $N_R$ by one and shift the energy by
\begin{equation}
\Delta_\gamma=\frac{e^2N_{SET}}{C_\Sigma}\frac{C_{RS}}{C_{\Sigma_R}}
\end{equation}
It is noteworthy that $\Delta_\gamma=\Delta_\alpha+\Delta_\beta$ which confirms that the effect of simultaneous tunneling can be evaluated as the combination of the effects of the two single tunneling events. A generic inter-dot coupling mismatch would suffice to produce discrete energy shifts of different magnitude and direction. Indeed, in the typical experimental condition where $C_{LS}>C_{RS}$ and $C_{\Sigma_R}\approx C_{\Sigma_L}$, then $\Delta_\alpha>\Delta_\gamma>0$ and $\Delta_\beta<0$. In particular, if $C_{LS}=2C_{RS}$, then $\Delta_\alpha=2\Delta_\gamma$ and $\Delta_\beta=-\Delta_\gamma$, as sketched in Fig.~\ref{fig:levels}(b). This would be directly reflected in the detector's current response and accounts for all the observed features. \\\indent
In conclusion, we have demonstrated charge sensing of a Si:P DQD connected to an electron reservoir by using a SET as an electrometer. Characteristic features in the detector's response allow one to identify the nature of the electron transfer and the number of excess electrons on each dot. Charging and detection mechanisms have been studied by means of low temperature measurements in conjunction with a classical capacitance model. The very precise charge density control demonstrated here can be useful in the future to operate these devices in the few electron regime, of interest for quantum information processing.\\\indent
This work was partly supported by Special Coordination Funds for Promoting Science and Technology in Japan.
\begin{figure}[t]
\includegraphics[scale=0.65]{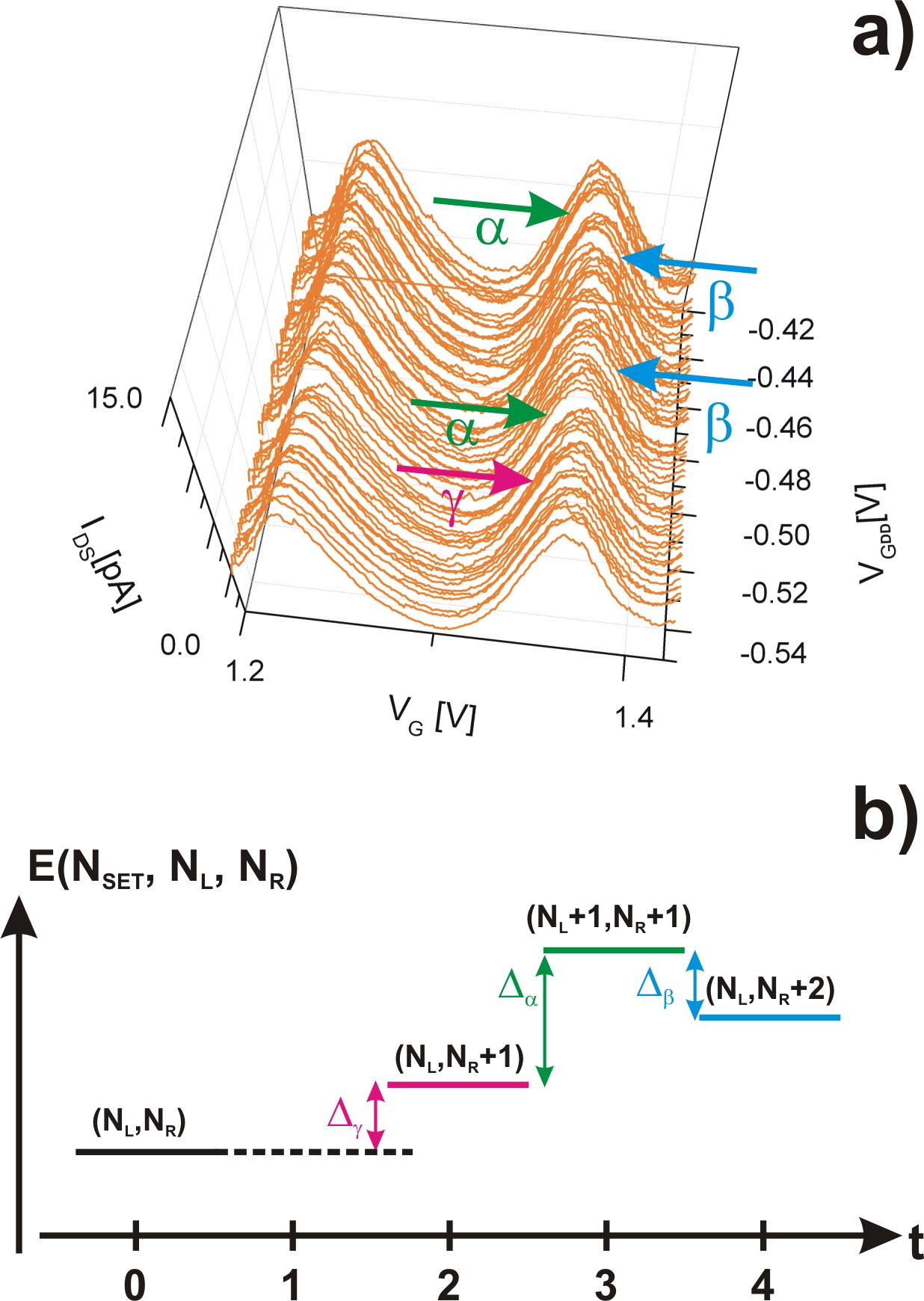}
 \caption{\label{fig:levels}(Color online) a) SET CB oscillations for different DQD gate voltages. The peak's position is shifted in $V_G$ whenever a single charging event in the DQD is triggered by an appropriate value of $V_{G_{DD}}$. Some shifts are highlighted by arrows (colors and lettering as in Fig.~\ref{fig:data}). Compensation is attained by simultaneously sweeping $V_{G_0}$. b) Electrostatic potential of the SET dot as a function of the number of tunneling events ($t$) occurring in the DQD. Excess number of electrons in either dot is reported for each transition. Magnitude and direction of the energy shifts derive from assuming $C_{LS}=2C_{RS}$ and $C_{\Sigma_R}\approx C_{\Sigma_L}$. }
 \end{figure}


%
%



\begin{thebibliography}{20}




\bibitem{fujisawa}
T.~Fujisawa, T.~Hayashi, and S.~Sasaki.
\newblock {\em Rep. Prog. Phys.}, 69:759, 2006.

\bibitem{angus}
S.J. Angus, A.J. Ferguson, A.S. Dzurak, and R.G. Clark.
\newblock {\em Nano Lett.}, 7:2051, 2007.

\bibitem{single}
C.~Single, F.~E. Prins, and D.~P Kern.
\newblock {\em Appl. Phys. Lett.}, 78:1421, 2001.

\bibitem{idqd}
E.~Emiroglu, D.~G. Hasko, and D.~A. Williams.
\newblock {\em Appl. Phys. Lett.}, 83:3942, 2003.

\bibitem{gorman}
J.~Gorman, D.~G. Hasko, and D.~A. Williams.
\newblock {\em Phys. Rev. Lett.}, 95:090502, 2005.

\bibitem{augke}
R.~Augke, W.~Eberhardt, C.~Single, F.~Prins, D.~Wharam, and D.~Kern.
\newblock {\em Appl. Phys. Lett.}, 76:2065, 2000.

\bibitem{mizuta}
M.~Manoharan, S.~Oda, and H.~Mizuta.
\newblock {\em Appl. Phys. Lett.}, 93:112107, 2008.

\bibitem{fab}
M.G. Tanner, D.~G. Hasko, and D.~A. Williams.
\newblock {\em Micro. Eng.}, 83:1818, 2006.

\bibitem{mypaper}
A.~Rossi and D.G. Hasko.
\newblock {\em Journ. Appl. Phys.}, 108:034509, 2010.

\bibitem{zimm_offset}
N.M. Zimmerman, W.H. Huber, B.~Simonds, E.~Hourdakis, A.~Fujiwara, Y.~Ono,
  Y.~Takahashi, H.~Inokawa, M.~Furlan, and M.W. Keller.
\newblock {\em Journ. Appl. Phys.}, 104:033710, 2008.

\bibitem{stone}
N.~Stone and H.~Ahmed.
\newblock {\em Appl. Phys. Lett.}, 77:744, 2000.

\bibitem{hofmann}
F.~Hofmann, T.~Heinzel, D.A. Wharam, J.P. Kotthaus, G.~B\"ohm, W.~Klein,
  G.~Tr\"ankle, and G.~Weimann.
\newblock {\em Phys. Rev. B}, 51:13872, 1995.

\bibitem{gareth}
G.J. Podd, S.J. Angus, D.A. Williams, and A.J. Ferguson.
\newblock {\em Appl. Phys. Lett.}, 96:082104, 2010.

\end{thebibliography}

\end{document}